# Towards Global Quantum Key Distribution

Haoran Zhang[1,2,6], Haotao Zhu[1,2,6], Ruihua He[2], Yan Zhang[2], Chao Ding[2], Lajos Hanzo[3,†] and Weibo Gao[1,2,4,5,†]

[1]School of Electrical and Electronic Engineering, Nanyang Technological University, Singapore, Singapore

[2]Division of Physics and Applied Physics, School of Physical and Mathematical Sciences, Nanyang Technological University, Singapore, Singapore

[3]School of Electronics and Computer Science, University of Southampton, Southampton, U.K.

[4]Centre for Quantum Technologies, Nanyang Technological University, Singapore, Singapore

[5]Quantum Science and Engineering Centre (QSec), Nanyang Technological University, Singapore, Singapore

[6]These authors contributed equally: Haoran Zhang, Haotao Zhu

†Email: hanzo@soton.ac.uk, wbgao@ntu.edu.sg

**Quantum Key Distribution (QKD) supports the negotiation and sharing of private keys with unconditional security between authorized parties. Over the years, theoretical advances and experimental demonstrations have successfully transitioned QKD from laboratory research to commercial applications. As QKD expands its reach globally, it encounters challenges such as performance limitations, cost, and practical security concerns. Nonetheless, innovations in satellite-based QKD and the development of new protocols are paving the way towards a worldwide network. In this review, we provide an overview of QKD implementations, with a focus on protocols, devices, and quantum channels. We discuss the challenges of practical QKD and explore long-haul QKD. Additionally, we highlight the future research directions that are expected to significantly advance the realization of a global QKD network.**

1. Introduction

   In an information-driven society, concerns about information security are paramount, where legitimate users face potential adversaries. This ongoing conflict, driven by complementary concerns about protecting information versus the desire to intercept it, can be quantified by comparing the computational resources and information access capabilities of both legitimate users and adversaries. Encryption remains the most common method for secure communication; however, the security of most cryptographic schemes relies on the limited computational power of adversaries, thus providing computational security rather than information-theoretical security[1]. The introduction of quantum mechanics imposes natural limitations on an adversary's ability to access information[2]. For instance, quantum wiretap channel theory[3] ensures that pre-shared quantum states can be utilized[4] to provide information-theoretical security. Within the field of quantum cryptography, quantum key distribution (QKD)[5] garners significant interest due to the intuitive role of securely shared keys in encryption. QKD enables key distribution between distant users at information-theoretical security and is proven to be composable[6] with further cryptographic schemes, such as the one-time pad or advanced encryption standard. Thus, it provides a versatile option for encryption in classical communication.

   Classical networks are already in use worldwide, enabling seamless communication between users. Therefore, a global QKD network, where users can securely distribute keys with one another, is always in demand to protect the overall security of classical networks. While not all communications may require quantum-secure encryption, it is crucial to ensure that the backbone of the network is covered. Therefore, as QKD applications continue to evolve, establishing a global QKD network that extensively links users around the world remains the ultimate goal in commercial applications.

   On the other hand, QKD faces questions regarding its practical value and reliability. The primary challenge comes from post-quantum cryptography (PQC)[7], which claims to be secure against quantum computers. While QKD maintains its unique advantage by offering information-theoretical security, something post-quantum cryptography as a form of classical cryptography cannot provide, the latter still alleviates public concerns about the threat posed by quantum computers[8] to information security. Another concern is QKD's inherent vulnerability to denial-of-service (DoS) attacks. Unlike classical DoS attacks, those targeting QKD networks occur at the physical layer, which is easier to monitor[9]. However, these concerns have already contributed to negative perceptions that the QKD community must address to reassure the public[10].

As QKD technology advances, the maturity of its various components varies significantly. Some aspects, such as Bannett-Brassard 1984 (BB84)-like protocols[5], have moved beyond the science-driven phase and are being explored for commercial applications. In contrast, the quantum memory network[11] —the envisioned backbone for all QKD applications—remains in its early stages. Consequently, predicting the optimal design of a QKD network is premature, as many related technologies continue to face significant challenges. Nevertheless, by examining and understanding these challenges, we can gain insight into the cutting edge of QKD research and anticipate future developments with greater clarity.

This review commences in section 2 by introducing the fundamentals of QKD implementation, covering the classification of different protocols, the key quantum devices and their operation, and the options for establishing quantum channel links. Then section 3 discusses the main challenges facing QKD applications, including key rate limitations, the cost-performance trade-off, and security concerns related to practical implementations. While confronting these challenges, several approaches in section 4 are being explored for achieving long-distance QKD, such as trusted relays, protocols that surpass theoretical bounds, and satellite-based QKD. The review discusses the advantages and drawbacks of these methods through examples of current applications. These efforts aim for extending the range of QKD are foundational for building a global quantum communication network. Finally, section 5 outlines the current developments in both terrestrial and satellite-based QKD networks. By examining the challenges and applications discussed, section 6 provides an outlook for the future research directions aiming for achieving a global QKD network. This review serves as both an introduction and an inspiration for fostering collaboration among academia, government, and industry within the field.

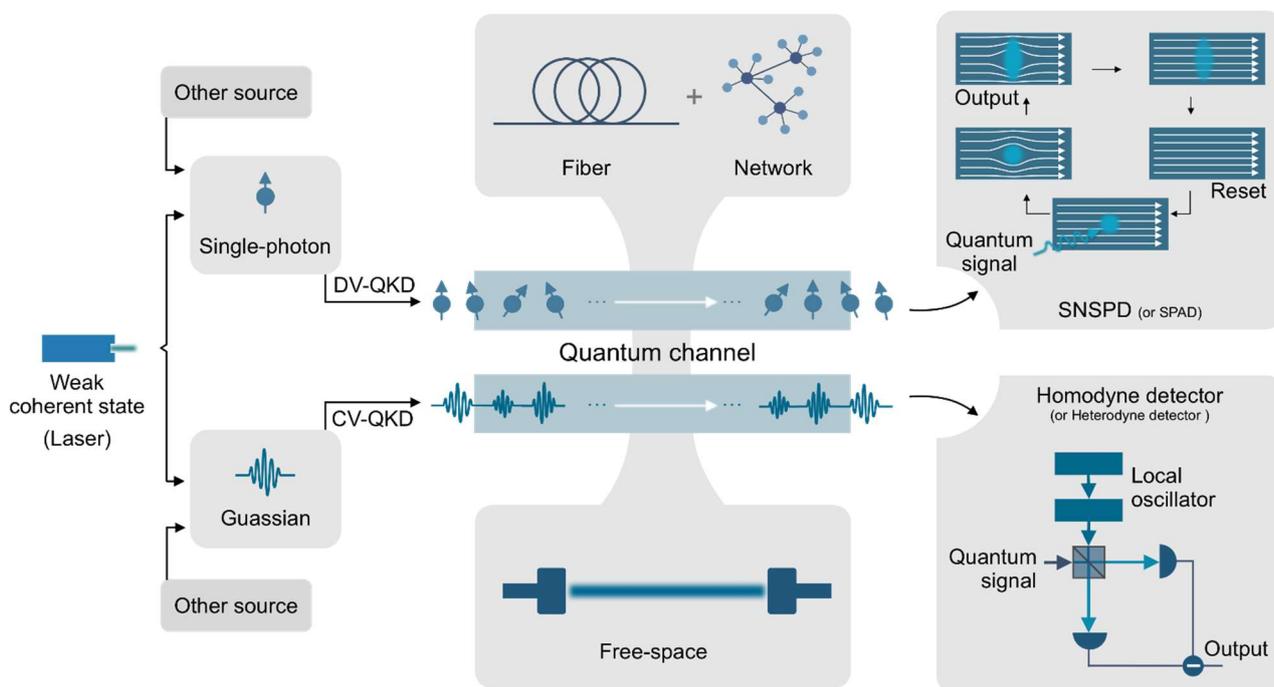

**Figure 1** Schematic of QKD systems employing DV single-photon state encoding and CV quadrature field amplitude encoding. While both typically utilize weak coherent state sources, the detectors operate on fundamentally different principles. The quantum channel is commonly implemented via fiber or free-space links.

2. Implementation of QKD

   a) Protocols

   QKD protocols are primarily categorized into two types: discrete-variable (DV) and continuous-variable (CV) protocols, as shown in Figure 1. In DV-QKD, information is mapped, for example, onto single photons or weak coherent pulses[5] in a discrete manner. These are subsequently detected using single-photon detectors[12]. DV-QKD protocols utilize quantum systems that are described by finite-dimensional Hilbert spaces, including states of polarization[5], phase[12], or orbital angular momentum (OAM)[13]. From this perspective, protocols such as differential-phase-shift (DPS)[14], coherent-one-way (COW)[15] and Round-Robin DPS[16] are also recognized as DV protocols. These distributed-phase-reference protocols are sometimes listed separately[17] due to the unique challenges they present in security analysis[18].

Conversely, CV-QKD[19] encodes information[20] onto the field quadrature amplitudes of quantized electromagnetic fields, which are detected using homodyne or heterodyne[21] detectors. The primary CV-QKD protocols employ coherent[22,23] or squeezed[24] Gaussian states encoding with Gaussian[25] or discrete keys within infinite-dimensional Hilbert spaces[26]. Additionally, protocols that utilize thermal states[27], discrete modulation[28] or unidimensional preparation[29] are also classified as CV protocols.

b) Devices

The implementation of QKD necessitates the use of specialized quantum devices, primarily consisting of a light source, modulators, and detectors. Many of these devices are similar to those used in classical optical communications. For the light source, while a single photon source is ideal for many DV-QKD protocols, weak coherent pulses are more commonly employed to simulate single photon sources due to their robustness and efficiency. For CV-QKD, the measurements require the homodyne or heterodyne detectors. These detectors, where the signal interferes with a local oscillator (LO), are not exclusively designed for quantum procedures. In contrast, for DV-QKD, the most indispensable devices are single photon detectors, such as single-photon avalanche diodes (SPADs)[30,31] and superconducting nanowire single-photon detectors (SNSPDs)[32]. Single photon detectors can precisely provide both the presence and timing of a single photon arriving along the optical paths that connects to the detectors. Assisted by the optical design at the receiver, where the measuring states are transformed into different paths, these detectors effectively perform the required measurements.

For near-infrared SPADs, the avalanche photodiode typically uses an indium gallium arsenide (InGaAs) layer for absorption and either an indium phosphide (InP) or indium aluminum arsenide (InAlAs) layer for multiplication[33]. Unlike operating in linear-mode, SPADs work in Geiger-mode[34], where the absorption of a single photon can initiate the impact-ionization process. This leads to a macroscopic, self-sustaining avalanche current that is detectable[34]. However, the electrical avalanche amplification structure is highly sensitive to environmental noise, and even noise from the avalanche process itself, resulting in a higher dark count rate (DCR)[35] and increased probability of afterpulsing[36]. To address this, a carefully designed quenching circuit[37] is essential to rapidly suppress the avalanche and reset the SPAD to its initial bias condition, although this introduces additional dead time as a trade-off. Optimizing the design of the avalanche photodiode structure, such as increasing the multiplication region thickness[38], can reduce the DCR. However, this may further compromise photon detection efficiency (PDE).

The performance of SNSPD surpasses that of SPAD across most metrics. Notably, SNSPD does not require a quenching circuit to suppress avalanche, which significantly shortens the reset time and enables a higher maximum count rate. Furthermore, SNSPDs demonstrate superior PDE and DCR performance, particularly in the near-infrared spectrum. For example, the typical PDE of InGaAs SPAD is around 30%, whereas SNSPD can achieve a maximum PDE of up to 99.5%[39]. Similarly, while an SPAD typically exhibits a DCR around several tens of thousands of Hertz, SNSPDs can reduce this to as low as 0.1 Hertz[40].

The operation of SNSPD is based on a mechanism where a supercurrent assists in the formation of non-superconducting regions, as shown in Figure 1, enhancing PDE, time precision and noise robustness as well. Typically, the nanowire, made of materials such as niobium nitride (NbN)[41], niobium titanium nitride (NbTiN)[39], tungsten silicide (WSi)[42], or molybdenum silicide (MoSi)[40,43], is cooled to well below its superconducting transition temperature and biased with a constant current near its critical value. In this superconducting state, Cooper pairs—charge carriers formed through electron-phonon interactions—are stable under Bose-Einstein condensation[44]. Upon photon absorption, if the photon's energy surpasses the binding energy of a Cooper pair, it breaks into two quasiparticles. For example, a 1550 nm photon can break 125 Cooper pairs in a NbN superconductor[44]. The generation of hundreds of quasi-particles forms a non-superconducting 'hotspot.' Subsequently, increased current density and vortex-assisted mechanisms[45] cause the hotspot to expand, eventually encompassing the entire cross-section of the nanowire. This expansion leads to an instantaneous increase in resistance, which is then detected.

c)  Quantum channel

In the general settings of QKD protocols, besides authenticated classical channels, quantum channels are required for sharing quantum resources between legitimate users. Quantum channels establish a physical connection between QKD users, allowing quantum states to be transmitted at the lowest possible environmental perturbation. Optical fibers and free space links are the most commonly used quantum channels. Single-mode fibers (SMF) are the preferred choice, as modern telecommunications networks are already extensively built on SMF operating at telecommunication wavelengths. Since single photons cannot be amplified like classical signals, the low-loss and stable transmission characteristics of telecom-band SMF make them ideal for quantum state transmission. Although issues such as birefringence, scattering[46], group-velocity dispersion, and polarization mode dispersion still exist, even in underground dark fibers, their effects can be substantially mitigated[47]. As a result, a QKD field test over 500 km has been successfully achieved without the assistance of any trusted nodes[48]. The establishment of a free-space quantum channel offers greater flexibility than fiber optics, especially when navigating challenging terrains or setting up temporary links. Additionally, free-space channels can cover much longer distances as the attenuation in free space is significantly less than in fiber. However, additional noise factors such as stray light and turbulence need to be addressed. Stray light can cause serious background noise, while turbulence can lead to space-time redistribution as well as beam spreading and wandering. Consequently, the use of spectral, spatial, and temporal filters, along with careful selection of operating times, spaces, wavelengths, and quantum basis, are essential considerations[49-51]. For instance, a metropolitan entanglement-based free-space network[52] can be established using a tall building as the central node. For unmanned aerial vehicle[53] or satellite QKD[54], developing a fast optical tracking system is also crucial for maintaining a stable and low-loss quantum channel[55]. Additionally, it is worth noting that the quantum channel for terahertz

QKD can be readily established using wireless technologies, albeit with a limited communication range of approximately a hundred meters[56].

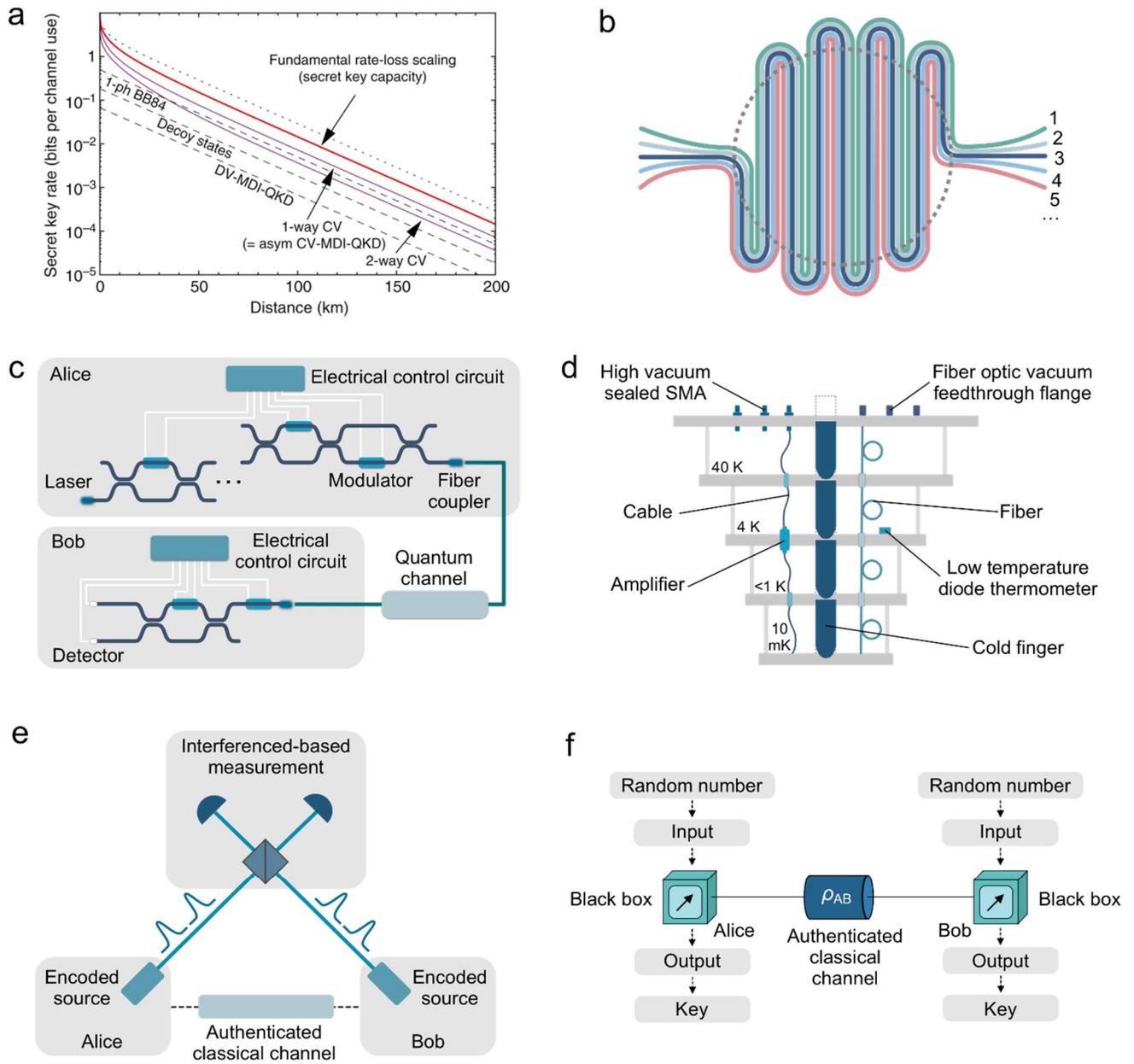

**Figure 2** (a) Simulation of ideal performances for different QKD schemes.[57] (b) Schematic of SNSPD chip with multiple interleaved pixels. (c) Schematic of on-chip QKD systems. (d) Schematic of multi-stage cryogenic cooling system. (e) Setup of MDI-QKD system. (f) Schematic of DI-QKD protocol.

## 3. Practical challenges of QKD

### a) Key rate

QKD supports key agreement for encryption purposes. Once the key has been negotiated, the system operates as in classical encryption. Explicitly, the information-theoretical security can be achieved using one-time pad[1], where the key must be as long as the data sequence. However, the key generation rate of practical QKD systems is inherently limited, presenting significant challenges for encrypting large volumes of classical data transmission. Therefore, much research has focused on understanding and overcoming the constraints on QKD capacity.

#### i. Channel attenuation

Channel loss occurs when the photons flying in the quantum channel have a probability of disappearing, rendering them undetectable. This phenomenon can be described as the quantum states splitting into the environment. Specifically, despite the efforts of manufacturers invested into purifying the glass-core fiber to minimize absorption, intrinsic inhomogeneities still cause Rayleigh scattering, leading to optical attenuation throughout the fiber. In the absence of a quantum repeater, it is intuitively clear that the probability of photon detection will decay with the accumulation of channel loss. This significantly limits the operational distance of QKD networks, which must avoid amplification.

For point-to-point QKD over a pure-loss channel with a transmittance of $\eta$, the Pirandola–Laurenza–Ottaviani–Banchi (PLOB) bound[57] provides a tight limit on the secret key capacity. This bound, illustrated in Figure 2(a), demonstrates that the secret key rate per channel use is constrained to $-log_2(1-\eta)$, approximating a near linear scale of $1.44\eta$ at low transmittance—hence, it is sometimes referred to as the linear bound. This bound is achievable in CV-QKD with the use of quantum memory. In practical scenarios operating without quantum memory, the secret key capacity of CV-QKD is reduced to $0.72\eta$, which is lower than that typically achieved by DV-QKD.

Moreover, the reduced tolerance for channel loss in CV-QKD, combined with challenges in the post-processing procedure, leads to the more frequent application of DV-QKD in metropolitan QKD networks.

Consequently, developing new protocols[58] that can mitigate the limitations of secret key capacity imposed by channel loss, and constructing quantum channels with lower attenuation —such as free-space links and hollow-core fibers[59] —are critical for establishing a global QKD network.

ii. QKD signaling rate

When the secret key capacity per channel use is fixed, the key rate of a QKD system is typically proportional to the relative frequency of channel uses. However, this repetition rate is limited by the speed of optical and electrical modulation and demodulation. Specifically, device bandwidth limitations can affect the accuracy of state preparation and measurement, while time jitter from the source, SNSPD, and time tagger can lead to crosstalk between adjacent signals, all contributing to an increase in QBER. Implementations of DPS-QKD have reached speeds up to 10 GHz relatively early[60], as the issue of crosstalk is less severe in this context. Nonetheless, protocols like DPS-QKD[14] and COW-QKD[15] have encountered challenges regarding security vulnerabilities under general attacks[61]. In fact, the secret key capacity of COW-QKD has been proven to scale as $O(\eta^2)$ [18,62], suggesting that, despite their widespread use, these original protocols may not be suitable for QKD networks requiring stringent security. For one-way BB84-like protocols, QKD systems have achieved repetition rates of up to 5 GHz for polarization encoding[63] and 2.5 GHz for time-bin and phase encoding[64].

To further increase the repetition rate, enhancing device performance[65], designing ultrafast optical modulation schemes[66], or developing passive protocols[67] would be crucial. However, since the key rate only increases linearly with the repetition rate, and since repetition rates have already reached a bottleneck at the GHz level, the QKD community is now focusing less on further improvements in this area.

iii. Saturation of detector

Assuming that low-loss channels and high repetition rates are achieved, the saturation of single-photon detectors in DV-QKD could still limit the key rate of the system[68]. This saturation arises due to the dead time, during which SPADs or SNSPDs become temporarily insensitive to incoming photons as they reset to their initial state for subsequent detection. To address this, multi-pixel SNSPDs have been developed[69] to maximize the key rate, as illustrated in Figure 2(b). When one pixel detects a photon and begins resetting, the remaining pixels are still active and ready to detect additional photons. The multi-pixel design not only shortens the dead time by reducing the length of each pixel, but also reduces time jitter[70]. This enhancement improves the saturation count rate of SNSPDs and subsequently increases the overall key rate to 110 Mbps[71].

b) Cost

Scientific frontier research typically focuses on discovering brand new phenomena, achieving breakthroughs, or enhancing performance, rather than addressing the challenges of commercial applications. However, society remains concerned about the costs associated with adopting new technologies. When deploying a large-scale QKD network, it is crucial to strike a compelling balance between benefits and costs.

i. Channel

Establishing stable quantum links is fundamental for constructing QKD networks. However, regardless whether using free-space or fiber links, extensive resources and close cooperation with governmental bodies are required. For instance, satellite links and other aerial vehicles are necessary, and even the ground stations are ideally situated on tall buildings or on plateau regions. Fortunately, the presence of dense classical networks provides potential fiber link resources for QKD network deployment, making it a cost-efficient option to share fibers with classical communication. However, the coexistence of quantum signals with classical optical communications demands effective isolation of the quantum channel from noise[72]. Using a dedicated frequency band[55], along with high-isolation wavelength division multiplexing[73] devices or optical filters[74], can help mitigate environmental disturbances. Even so, in-band noise caused primarily by Raman scattering[75] from high-power classical signals[76] must also be addressed[77] to ensure the efficient use of bandwidth resources.

However, certain devices commonly used in classical networks, such as reconfigurable optical add-drop multiplexers and optical communication repeaters[78], are not suitable for quantum signals. Instead, the quantum channel must bypass these devices. Furthermore, while classical optical communication can transmit over hundreds of kilometers of fiber with the help of repeaters, quantum signals over such long distances become too weak relative to the background noise[74]. Consequently, a "plug-and-play" integration of a QKD network into existing classical networks is not readily feasible. Significant efforts are required for designing the appropriate QKD network infrastructure[79] and coordinate with telecommunications companies before such integration can be realized.

ii. Form-factor

For widespread applications of QKD networks, quantum devices are expected to achieve a similar level of miniaturization as those in classical networks. Dense integration of quantum devices can enhance functionality and enable the construction of scalable QKD systems. In fact, QKD systems require precise modulation of quantum states, and lithographic precision makes this achievable even in mass production. Moreover, integrated structures make components like Mach–Zehnder interferometers robust against environmental noise[80], reducing the need for feedback systems[47].

Much of the research on on-chip QKD is based on silicon and silica photonics[81] due to its mature manufacturing techniques. Modulation on silicon chips primarily relies on thermo-optic effects, which limit the choice of QKD protocols[82]. However, carrier depletion in silicon enables active modulation for high-speed QKD[71]. Additionally, developments in on-chip lithium niobate[83] provide an alternative by facilitating easier electro-optic modulation.

The main advantages of chip-based QKD systems lie in achieving monolithic hybrid integration as illustrated in Figure 2(c). For QKD transmitters, it is crucial to integrate the light source directly onto the chip[84], making materials like InP[85] preferable for this purpose. For QKD receivers, integrating SNSPDs[86], SPADs[87], and homodyne detectors[88,89] onto the chip is essential. Since the benefits of integration become economically significant with mass production, the recent developments indicating that the QKD community is becoming increasingly prepared for the worldwide commercial deployment of QKD networks.

iii. Temperature

Single-photon detectors operating in the telecommunication wavelength range require significantly lower temperatures than classical detectors to minimize noise. For example, SNSPDs achieve optimal performance only at temperatures much below half the critical temperature of their superconducting material —practically around 2–4 K for Nb(Ti)N-based SNSPDs[90]. This necessitates multi-stage cooling systems as illustrated in Figure 2(d), such as Gifford–McMahon[90] or Joule–Thomson cryocoolers[91], which have considerable size, weight, and power requirements when implemented at base stations within classical networks. Although there is ongoing research to develop SNSPDs from materials with higher critical temperatures[92], the fundamental operating principle of SNSPDs implies that the cost of cryogenic cooling systems cannot be entirely eliminated. Centralizing the SNSPD chips could reduce the number of cryogenic systems needed, making this design consideration crucial for the overall structure of QKD networks.

In contrast, SPADs can operate near or even at room temperature[93], although they generally suffer from lower detection efficiency and higher noise levels. A typical thermoelectric cooler provides adequate cooling for SPADs, avoiding the need for cryogenic systems and offering an economical and flexible option for equipping QKD receivers.

While SNSPDs are eminently suitable for QKD applications in terms of performance, the cost implications for commercial use remain a significant consideration. From this perspective, advances in cost-effective cryocoolers will play a crucial role in shaping the future infrastructure of QKD networks.

c) Security

QKD promises to guard against all types of attacks encountered in quantum mechanics and information theory, thereby providing information-theoretic security for shared keys. However, despite the rigorous derivations in security analysis, it generally assumes that all user devices are perfectly modeled—an assumption that rarely holds in practice. Such discrepancies could lead to additional unexpected information leakage beyond the bounds set by the security analysis, a phenomenon known as "side-channels." These side-channels pose significant challenges for quantum protocols, particularly because they involve highly sensitive quantum devices designed to ensure rigorous security. Therefore, addressing these side-channel vulnerabilities is crucial for the security of QKD applications.

**Table 1. Summary of quantum attacks against QKD**

| Quantum attack | Target device | Protocol | Countermeasures |
| --- | --- | --- | --- |
| Calibration attack[94,95] | Local oscillator | CV-QKD | Real-time shot noise measurement[94] and hidden-Markov-model-based recognition[95] |
| Saturation attack[96] | Homodyne detector | CV-QKD | Gaussian post-selection |
| Wavelength attack[97] | Beam splitter | CV-QKD | Incorporating a simple wavelength filter |
| Laser-seeding attack[98] | Source | DV-QKD | Incorporating an external isolator |

| Source attack[99] | Source | DV-QKD | Implementing phase randomization |
| Phase-remapping attack[100,101] | Phase modulator | DV-QKD | Enhancing timing control and implementing phase randomization |
| Blinding attack[102,103] | Homodyne detector[102] Avalanche photodiode[103] | CV-QKD[102] DV-QKD[103] | Using a sensitive p-i-n photodiode[102] and a separate watchdog detector[104] |
| LO-intensity attack[105] | Local oscillator | CV-QKD | Monitoring the LO intensity[105] and machine-learning-based detection[106] |

i. Attack on devices

A practical QKD system may be vulnerable to various quantum attacks due to the practical imperfections of devices. The typical quantum attacks against practical QKD systems are listed in Table 2, such as calibration attacks[94,95], saturation attacks[96], wavelength attacks[97], laser-seeding attacks[98], source attacks[99], phase-remapping attacks[100,101], blinding attacks[102,103], and local oscillator (LO)-intensity attacks[105]. Fortunately, in response to the challenges posed by these quantum attacks, effective countermeasures have been proposed. These can be broadly categorized into two main types: incorporating suitable monitoring modules and integrating machine-learning detection modules. In monitoring modules, leveraging suitable monitoring devices to track the physical parameters of pulses—such as intensity, phase, and wavelength—is a countermeasure to guard against specific quantum attacks. However, these monitoring devices can introduce potential security vulnerabilities due to inherent imperfections and increase the overall complexity of QKD systems. In detection modules, employing machine learning algorithms to identify various patterns and features of quantum attacks serves as an effective countermeasure for achieving universal attack detection in specific protocols. However, these machine-learning detection modules can handle most quantum attacks on CV-QKD protocols, but they are unable to detect quantum attacks on other protocols. Therefore, it is essential to develop a universal countermeasure that can protect any practical QKD protocol from various quantum attacks.

ii. Measurement-device-independent QKD

In one-way quantum communication, attacks on the transmitter's devices are relatively less concerning, as they can be more easily monitored[107]. However, isolators are not suitable for the receiver, and the measurement of quantum states requires the receiver's devices to be highly efficient and sensitive. Consequently, many challenges in defending QKD systems against practical attacks are linked to their measurement devices. To address this issue, measurement-device-independent QKD (MDI-QKD) protocols[108] — also known as side-channel-free QKD[109] —were proposed in 2012. These protocols eliminate the need for specific assumptions about modeling the measurement devices, alleviating major concerns about side-channel attacks.

To elaborate a little further, for MDI-QKD protocols, two legitimate users each generate quantum states, while a third party positioned between them performs interference-based measurements, as illustrated in Figure 2(e). Although high-fidelity measurements are still required in MDI-QKD, they do not need to be trusted. The third party is treated similarly to the quantum channel, with security verified through protocol procedures, such as comparing a random sample of measurement results with the prepared states. It is worth noting that typical MDI-QKD protocols require two-photon interference with pre-selected measurements, resulting in roughly the same secret key capacity as point-to-point QKD, which scales as $O(\eta)$. While MDI-QKD resolves security concerns related to measurement devices, it introduces additional untrusted nodes and necessitates precise synchronization to account for timing differences between the two channels. Moreover, practical setups with asymmetric node positions can decrease secret key capacity[110]. As a result, MDI-QKD has not been as widely applied as point-to-point QKD for network establishment[111], until the development of new protocols that surpass the PLOB bound[58].

iii.  Device-independent QKD

Device-independent QKD (DI-QKD) protocols[112] take a step further by eliminating the need for any specific assumptions about the modeling of all quantum devices, aiming for achieving the ideal level of security promised by QKD[113]. In a device-independent scenario, legitimate users treat local quantum devices as 'black boxes', focusing solely on their classical inputs and outputs rather than on their internal operations. As a result, any potential flaws in the quantum devices—regardless whether due to intrinsic imperfections or malicious tampering—can, in principle, be eliminated. Specifically, DI-QKD distributes the secure key through nonlocal correlation[112], with the parameter estimation resembling the process of verifying the violation of Bell inequality[114], as illustrated in Figure 2(f). However, a key challenge remains in addressing the detection loophole: in DI-QKD, non-coincidence outcomes—where only one of the detectors clicks—must be accounted for, rather than discarded during post-selection, similar to the requirements of the loophole-free Bell test[115]. This makes photonic approaches[116] to DI-QKD challenging for long-distance applications, unless improved protocols[117,118] can relax the stringent detection efficiency requirements such as relying on a qubit amplifier[119] or remote Bell test[117]. In contrast, DI-QKD based on the establishment of distributed entanglement —such as systems using rubidium atoms[120] and trapped ions[121] —can circumvent the detection loophole through single-shot readout, offering a promising solution for DI-QKD networks.

However, can the DI-QKD community claim that their experiments are entirely "loophole-free"? Unlike in a loophole-free Bell test, where the locality loophole requires measurements to be space-like separated to exclude hidden-variable influences, this concern is not as critical in DI-QKD. The general settings of DI-QKD inherently close the locality loophole: under the assumption that quantum theory is correct and that no unintended information, such as the input and output, leaks from the users' ends[120].

Nevertheless, certain assumptions about the devices —such as the perfect unidirectional isolation[121] of the users' ends —still persist, leaving potential loopholes. Although the reliability of isolators and other classical devices may fall outside the scope of QKD research, as DI-QKD becomes widely applied in networks, these issues could still raise questions and concerns within society.

## 4. Long-haul QKD

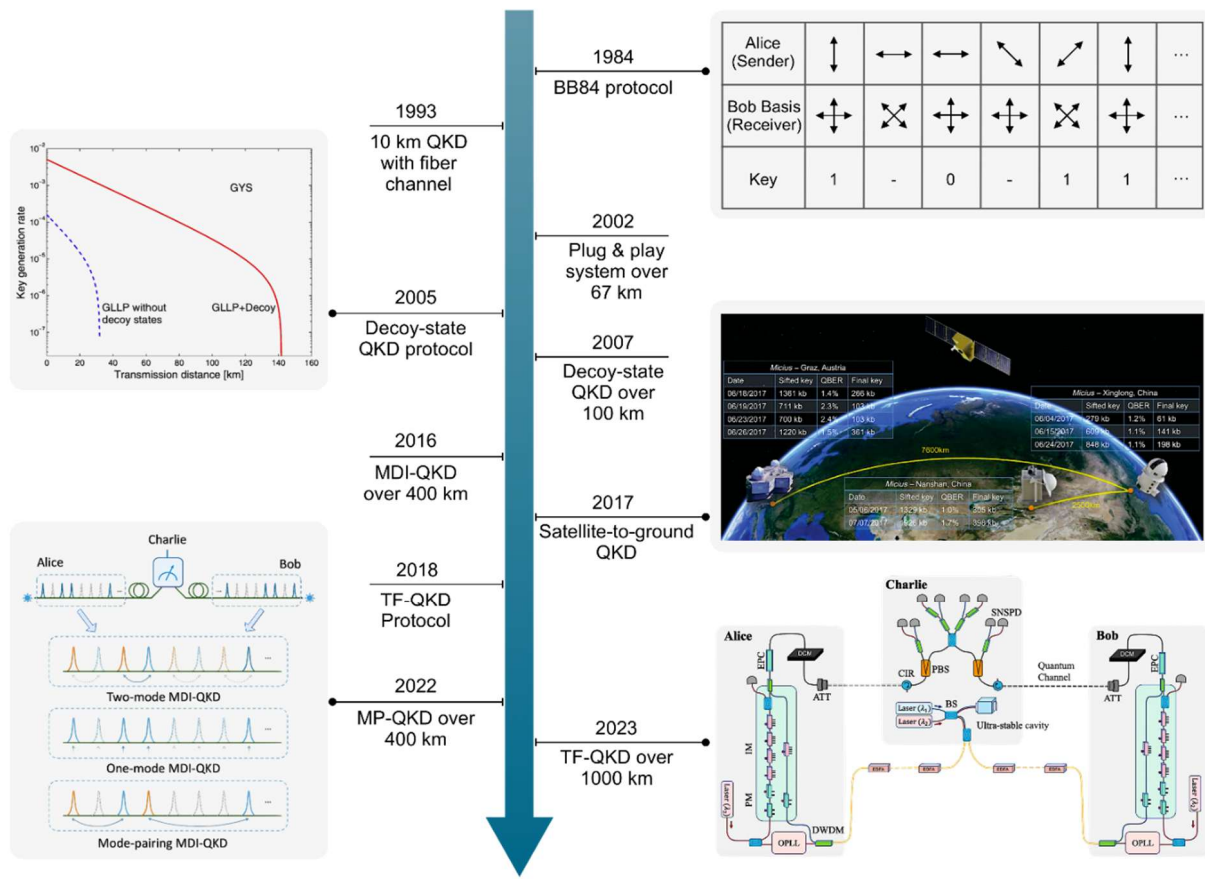

**Figure 3** Long-haul QKD developments. Since the proposal of the first QKD protocol, namely the Bennet-Brassard BB84 protocol in 1984, the field of QKD has received widespread attention. In the initial stages, the range of QKD systems were significantly limited. To consider their evolution briefly, in 1993, a secure distance of 10 kilometers was achieved with the BB84 protocol[122]. In 2002, the plug-and-play system was designed, achieving a secure distance of 67 kilometers[123]. With the aid of the decoy state protocol, the practicality of QKD was elevated to a new level[124-126]. The secure distance of QKD exceeded 100 kilometers[127] using the decoy state method[124]. In 2009, QKD over 250 kilometers[128] was achieved by utilizing ultra-low-loss optical fiber. In 2012, the MDI-QKD protocol was proposed[108],

which is of great significance in quantum networks given its capability of closing all measurement-side loopholes. This protocol has driven the development of long-haul QKD. In 2016, MDI-QKD over a fiber distance of 404 kilometers[129] was accomplished, breaking the record for fiber-based QKD at that time. In 2017, the first satellite-to-ground QKD[54] and satellite-relayed intercontinental quantum network[130] were achieved. In 2018, the proposal of the TF-QKD protocol improved the relationship between the secure key rate and transmission rate from a linear to a square root shaped function but also surpassed the theoretical upper limit of point-to-point key rate, enabling longer distances and higher rates of QKD[58]. In 2020, using ultra-low-loss optical fiber, TF-QKD over a distance of 509 kilometers[131] was achieved. In 2022, QKD over 830 kilometers of optical fiber was demonstrated using the TF-QKD protocol[132]. The MP-QKD protocol without global phase locking was proposed[133] in the same year. In 2023, the distance of TF-QKD experiment exceeded 1000 kilometers[134].

Long-haul QKD serves as the backbone for establishing connectivity in global quantum networks. In recent years, researchers have made substantial efforts in this area, leading to numerous significant advances, as depicted in Figure 3. Considering the trade-off between key rate and channel loss, various approaches have been proposed and explored for implementing long-haul QKD.

a) Trusted relay

Similar to classical communication, the light intensity of quantum communication also decreases as the communication distance increases. Therefore, to extend the transmission distance and coverage area of quantum networks, the most straightforward solution for long-haul QKD is to increase the number of relays in the quantum network[135]. Hence, there has been extensive research on quantum repeaters, even in the real world[136], which can extend communication distances without the need for trust. However, their application in practical quantum networks would incur significant costs, so we will not discuss fully-fledged entanglement-based quantum repeaters as a relaying method here. But the premise of trusted relays is that the relays must in protected customer promises that are inaccessible by eavesdroppers. In the real world, trusted relays between two cities are usually set up in suburban areas, so deploying real-time monitoring to check for eavesdropping will consume significant human and material resources.

The SECOQC quantum network[137] is an early example of using quantum relays to implement a quantum network. Subsequently, trusted relays have been used in many quantum networks. In Fig. 4, we show some networks that use quantum relays. How to deploy nodes as trusted relays in a multi-user quantum network to optimize the network's efficiency and the number of relays is also an important issue. There are many models and algorithms regarding this aspect[138]. Besides the trusted relay, in 2012, the Lo research group proposed MDI-QKD[108], which closed all vulnerabilities at the measurement side, hence dispensing with the requirement of trusted relays for measurement parts. This significantly enhanced the security of quantum networks. More importantly, if MDI-QKD is widely applied, the number of trusted relays can be significantly reduced, mitigating the practical costs. Rationally utilizing both trusted and untrusted relays to maximize the efficiency of quantum networks in terms of cost and performance is an essential consideration for the future development of a global quantum network[139].

b) Hybrid quantum-safe relay

PQC is widely considered as a compelling quantum-safe alternative to QKD, as it can provide computational security against quantum computers[7]. However, combining PQC with quantum communication can enhance security, with each method mutually reinforcing the other. Importantly, PQC can effectively mitigate security concerns associated with trusted relays in quantum networks, a concept referred to as a secure relay[140] or hybrid quantum-safe cryptosystem[141].

To secure trusted relays across quantum networks, one intrinsic solution involves transmitting ciphertext using quantum secure direct communication[142]. In this method, only information pre-encrypted by PQC is encoded onto quantum states and transmitted through quantum networks. This guarantees that, even if any network relay is compromised, the security of all encrypted messages remains quantum-safe. Alternatively, PQC can be integrated with the post-processing procedure of QKD for key derivation[141]. Furthermore, PQC can play a vital role in public key generation and authentication processes[143], establishing a robust and reliable classical infrastructure to support quantum networks.

c) Beating the repeater-less bound

Proposed in 2016, the PLOB bound was initially considered the definitive bound for all repeater-less QKD systems. However, it was subsequently recognized as inapplicable to interference-based QKD. In 2018, the research group from Toshiba's Cambridge Research Laboratory proposed the twin-field QKD (TF-QKD) protocol[58]. Unlike typical MDI-QKD protocols that rely on two-photon interference, this protocol is based on the principle of single-photon interference, allowing effective events from single-photon responses to be directly used for key generation. Additionally, as the TF-QKD protocol falls under MDI-QKD, it strategically positions the measurement device at the midpoint, effectively halving the distance a single photon travels. Consequently, this adjustment alters the key generation rate from being proportional to the transmission rate to being proportional to the square root of the transmission rate. In 2022, the mode-pairing QKD (MP-QKD) protocol, developed[133] and concurrently proposed as asynchronous MDI-QKD[144], demonstrated its ability to exceed the PLOB bound for key generation. The MP-QKD introduces a 'measure-then-pair' strategy based on MDI-QKD, significantly enhancing the utilization rate of response events. These protocols allow the key rate to be proportional to the square root of the transmittance as well, thus breaking the PLOB bound. Moreover, these protocols have been validated in laboratory settings and field fiber optic environments[48,131,145-152], achieving significant results and surpassing a communication distance of 1000 km[134]. However, despite surpassing the repeaterless bound, these single-interference based protocols still do not meet the single-repeater bound of $-\log_2(1-\sqrt{\eta})$ for end-to-end QKD[153]. Consequently, in the realm of global QKD, exploring strategies like increasing the number of ends performing interference-based measurement to surpass this new bound at the protocol level remains a crucial area for future research.

## d) Satellite QKD

Compared to the exponential decay of the photon transmission rate versus communication distance in optical fibers, the effective attenuation of photons in free space is related to the effective thickness of the atmosphere, with an approximate attenuation of 20 dB over 1000 km[154]. Therefore, using satellites is an effective method of supporting global QKD across countries and continents[155,156]. A European research group initially verified the feasibility of single-photon transmission via satellite-to-ground links in 2008[157]. Subsequently, in 2013, the Weinfurter group demonstrated the feasibility of QKD under high-speed motion using an aircraft platform[158]. A Chinese group also achieved quantum communication on a high-altitude balloon platform[159]. With various countries' increasing emphasis on quantum communication, such as the United States, Canada, the European Union, China, and Singapore, satellite-based quantum communication has rapidly developed. Currently, the largest project is the Micius Quantum Science Experimental Satellite launched by the Chinese Academy of Sciences[154]. Two ground stations connected by Micius, Xinglong and Nanshan, are 2600 km apart, and the link distance between the satellite and the ground stations can exceed 2000 km. This communication distance is currently the longest in single-relay QKD.

# 5. QKD network

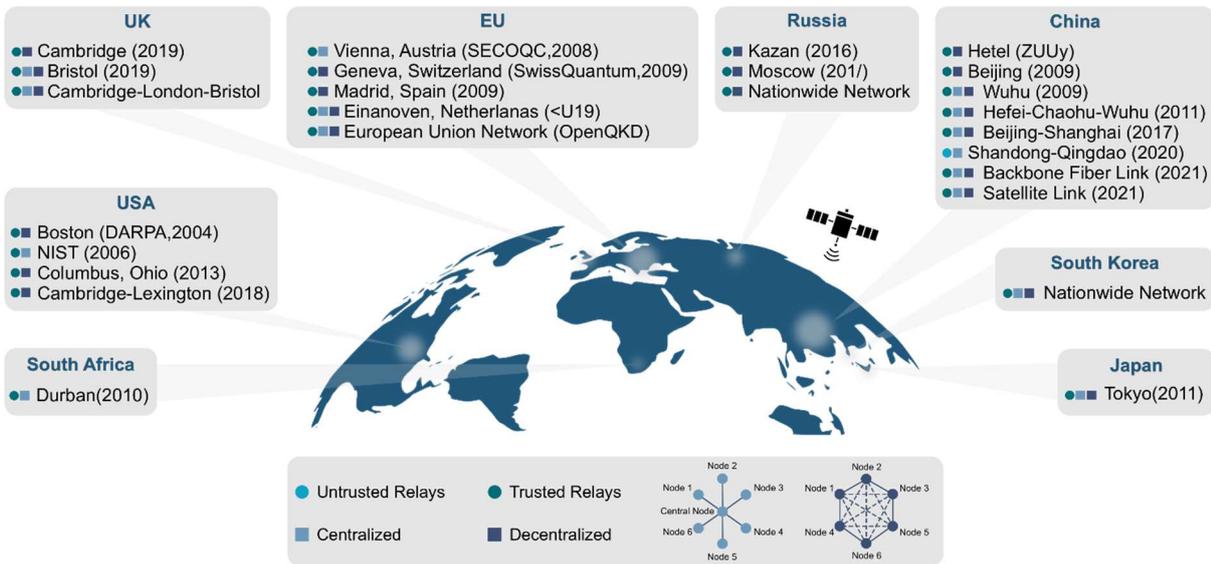

**Figure 4** Main existing QKD networks worldwide. As point-to-point QKD develops rapidly, the deployment of multi-user, multi-node QKD networks constitutes the next step forward. The first QKD network was established at three locations: Harvard University, Boston University, and the BBN Corporation. The number of nodes was later increased to ten[160]. Subsequently, a quantum network involving organizations and institutions from 12 countries including the UK, France, Germany, and Austria demonstrated its capabilities in Vienna[137]. Following this, China, Japan, and Switzerland all established quantum communication networks[111,161-166]. In 2021, China's quantum communication backbone network between Beijing and Shanghai relied on 109 nodes[166]. We have also labeled all the QKD examples in the figure according to the presence or absence of trusted relays and the topology structures covered in the quantum network.

As QKD research advances, especially in the current situation where there is a large number of users, deploying a multi-user, multi-node quantum network becomes inevitable. However, in the real world, there are many factors to consider when deploying a quantum network, such as the differing communication needs and distances between cities, the crosstalk noise between different channels connecting the servers and the users. For ground-based global quantum networks, optical fibers are typically chosen as the communication channel. While there are also experimental examples using free space, even water as channels[167,168], optical fibers are generally a better choice due to their resistance to electromagnetic interference and ease of deployment. Commercial communication fibers usually meet the requirements of quantum communication. Based on the communication range of quantum communication, terrestrial global QKD quantum networks can be divided into metropolitan and intercity quantum QKD networks.

a) Terrestrial QKD network

  i. Metropolitan area

  The metropolitan QKD quantum network usual involves a large number of nodes. For example, the quantum network in Hefei, a city in China's Anhui Province, has 46 nodes[169]. In a multi-node scenario, it is necessary to design a reasonable topology to optimize the overall key generation efficiency. At the same time, various practical factors in node deployment must also be considered, such as the fact that some locations may not be geographically suitable for being set up as intermediate trusted relays. In metropolitan area networks, the distance of optical fibers is usually only a few kilometers to tens of kilometers. A noteworthy point here is that in MDI-QKD, the measurement station is placed in the middle between the communicating parties. Since the measurement station requires single-photon detectors, multiple users often share the same detection node to minimize the number of intermediate detection nodes and reduce the cost. Therefore, in real-world scenarios, there may be cases where the distances from the communicating parties to the detection node are different, resulting in an asymmetric situation. There has been much research related to this issue to get higher key rate over the asymmetric situation[110,111,170,171]. One of the most straightforward ways to address an asymmetric path is to apply attenuation compensation to the less attenuated segment, making the situation symmetrical. However, researchers have optimized the parameters under asymmetric conditions and achieved a higher coding rate without needing attenuation compensation, surpassing the results obtained with added attenuation. Optical switching and secure key management in metropolitan quantum networks are also crucial and essential components of practical quantum network[111,163]. The switching can be categorized into two types: matrix switching and fully connected switching. Matrix switching does not allow any two users to connect freely, whereas fully connected switching enables users connected to the optical switch to establish connections between two endpoints.

ii. Inter-city area

For intercity quantum networks, the communication distance is typically on the scale of hundreds of kilometers. At such distances, it is essential to consider issues such as photon attenuation, dispersion, and scattering in the optical fiber[48]. In the real world, optical fibers spanning hundreds of kilometers usually have multiple splicing points, which can lead to problems like reflection and crosstalk at these splicing points[150,172]. Additionally, at the time of writing, most quantum networks still use protocols proportional to the square of the transmittance, meaning that any reduction in attenuation results in a squared decrease in the key generation rate. As shown in Figure 4, several countries, including the United States, Europe, China, and others, have already begun establishing intercity quantum networks. Moreover, in the current stage of intercity quantum networks, quantum information typically transmitted alongside classical optical communication[76], either in different optical cables or even in different fibers within the same cable. This often requires taking into account the various interferences introduced by classical optical signals and employing appropriate methods to reduce it. Especially in the case of large intercity distances, the impact of quantum noise is grave. Additionally, some classical synchronization signals often have to be amplified over intercity distances using optical amplifiers[150], making it even more important to pay attention to the crosstalk between the synchronization signals and optical quantum signals. Similar to metropolitan quantum networks, the setup of trusted relays in intercity networks also has to consider practical factors such as distance, fiber environment, and cost to ensure the efficient operation of the quantum network. It is important to emphasize that at metropolitan distances, the key rate of the BB84 protocol may be surpassed by the TF-QKD[58] or MP-QKD[133,144] protocols. However, at intercity distances, such as between two cities 200 km apart, the TF-QKD and MP-QKD protocols have demonstrated significant advantages regarding key rate and security. Therefore, TF-QKD and MP-

QKD protocols are likely to become a pair of important protocols for future intercity quantum networks. The diversity of implementation methods[173] for TF-QKD and the simplicity of MP-QKD experimental setups give these two protocols even broader application appeal.

b) Satellite-based QKD network

For a satellite-based quantum network, multiple satellites are required to complete seamless coverage of the system. Satellites can be typically divided into Low-Earth Orbit (LEO) satellites, Medium-Earthand Orbit (MEO) satellites, and Geostationary-Earth Orbit (GEO). The simplest satellite-based QKD network consists of a MEO satellite and three LEO satellites[154]. LEO satellites have shorter orbital periods, allowing them to provide 24-hour service, but they cover a smaller area so we need more of them for seamless coverage. On the other hand, GEO satellites have longer orbital periods, meaning that the effective time for QKD within a day is shorter, but they cover a larger area. Therefore, when building a satellite-based QKD network, both LEO, MEO and GEO satellites are needed to achieve efficient QKD. Besides, in actual satellite systems, the size and weight of the system are critical factors for the successful operation of the satellite[130]. Therefore, making the system more integrated is an important consideration in quantum satellite networks. It is anticipated that in the future, satellite-based quantum networks can better support global QKD.

6. Outlook

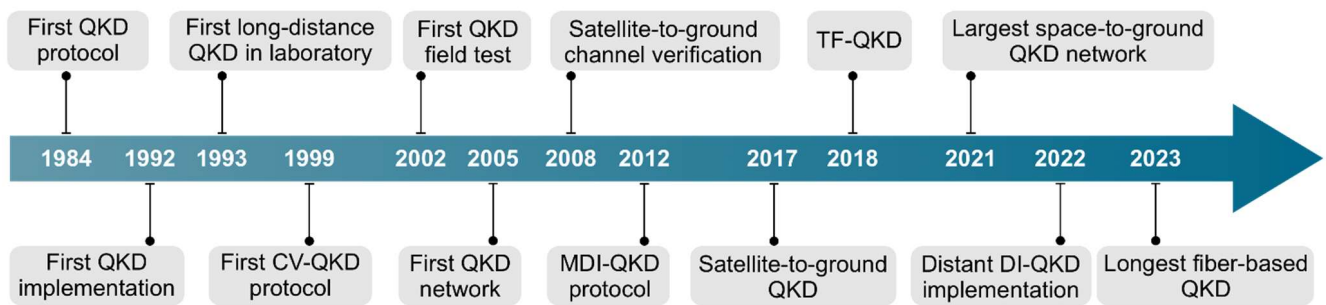

**Figure 5** Milestones towards global QKD networks.

In this review, we discussed the objectives and implementation methods of global QKD, as well as the challenges faced during the specific implementation process. We also highlighted the existing implementations of long-haul QKD and QKD networks. Figure 5 portrays the essential milestones in the progression from the inception of QKD to its realization in global networks.

This development includes significant achievements such as the proposal of the first QKD protocol[5], and its initial implementation[174], long-distance experiments[175], the introduction of the CV-QKD protocol[19], the first QKD field tests[123] and networks[160]. We also underline the evolution of satellite-based QKD networks from concept[157] to realization[54,166], the proposals of MDI-QKD[108] and TF-QKD[58] protocols alongside their longest fiber-based implementations[134], chip-based QKD[85], and the long-distance realization of DI-QKD[120]. These milestones collectively represent advances across several key directions: extending QKD transmission distances, increasing key rates, enhancing security, and integrating QKD systems.

For extending communication distances, the core issue is how to capture more of the encoded information carried by photons. In different scenarios, we choose lower-loss channels. For example, photon attenuation in the atmosphere is much lower than in optical fibers, hence longer distances may be attained (such as between two ground stations connected by Micius, Xinglong, and Nanshan, 2600 km apart[166]). Additionally, improving the detection efficiency of single-photon detectors is another crucial factor for extending communication distances. The PDE of SNSPDs[176] is higher than that of SPADs. Of course, protocols also play a crucial role. For instance, the TF-QKD[58] and MP-QKD[133] protocols make the key generation rate proportional to the square root of the photon transmission rate. In terms of security, thanks to the introduction of the decoy-state protocol, commercial weak coherent light sources can replace single-photon sources and be directly applied in the QKD field, making them immune to photon-number-splitting attacks, which has substantially promoted the practical application of QKD. The proposal of the MDI-QKD protocol has further mitigated the security vulnerabilities on the detection side. We have also noted that fully passive QKD protocols eliminate the side-channel security vulnerabilities introduced by active modulation[67,177]. In terms of device integration, a more integrated QKD system is important for large-scale user applications. For example, the newly launched microsatellite has a payload weight of about 23 kg, while the portable ground station weighs approximately 100 kg[178]. Additionally, chip-based QKD is an essential direction for device integration, and we look forward to the day when QKD will be widely accessible.

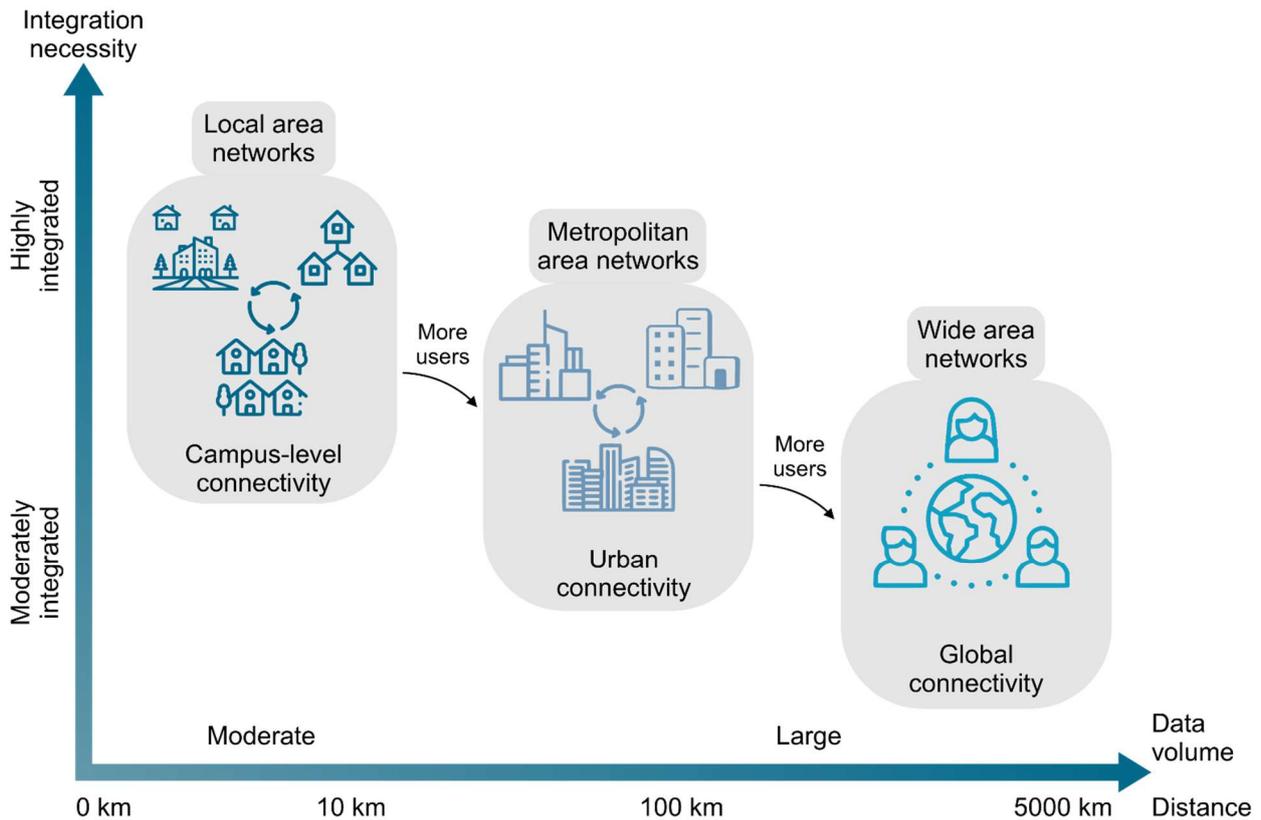

**Figure 6** An envisioned layout towards global QKD network. In terms of distance, we borrow the concepts and approximate range definitions of local area networks (LAN), metropolitan area networks (MAN), and wide area networks (WAN) from classical networking[179,180]. Typically, the number of users in a MAN is greater than that in a LAN, while a WAN has even more users than a MAN, which aligns with the user number comparisons found in these three types of networks in classical networking. Regarding integration requirements, LANs generally require a higher level of user system integration to meet user demands, while the integration requirements decrease for MANs and WANs. For LANs, an appropriately sized data center is usually needed to support user communication needs, whereas, in WANs, larger data centers and even larger satellites may be necessary to support QKD systems, which places higher demands on equipment such as light sources. Therefore, achieving higher integration is currently more challenging. The amount of communication information for users also increases progressively with the number of users. A common example of a LAN is a campus-level area, while a typical

example of a MAN is an urban network, and WANs encompass networks that exceed this range.

As a quantum counterpart to the internet[79,181], the global QKD network will consist of multiple layers, each tailored to specific application scenarios and requirements. Here, we outline the potential layout of our envisioned future global QKD network as a forward-looking perspective. Figure 6 illustrates our vision, which considers factors such as transmission distance, user numbers, integration needs, and data transmission volume across various application scenarios. When developing relevant technologies and designing practical QKD networks, it is crucial to thoroughly assess both the requirements and challenges involved. Notably, while key rate and cost are often prioritized, security issues in QKD projects tend to receive less attention from funders and contractors. This is primarily because the practical security level is challenging to evaluate quantitatively[182]. Additionally, efforts to monitor every potential vulnerability can increase costs and reduce performance. Current evaluation and testing methods[183], such as those published by the International Organization for Standardization and the International Electrotechnical Commission, can help identify typical device flaws.

Consequently, information security is essential for the orderly functioning of society, as it ensures that messages are accessible only to authorized parties, while remaining protected from unauthorized access. Global QKD promises to establish a worldwide, information-theoretically secure connection among all legitimate users. Society stands to benefit significantly from establishing a global QKD network offering practical capacity and cost-effectiveness.

**ACKNOWLEDGMENTS**

This work is supported by the National Research Foundation, Singapore and A*STAR under its Quantum Engineering Program (NRF2021-QEP2-01-P02, NRF2022-QEP2-02-P13), ASTAR (M21K2c0116, M24M8b0004), Singapore National Research foundation (NRF-CRP22-2019-0004, NRF2023-ITC004-001, NRF-CRP30-2023-0003, NRF-MSG-2023-0002), Singapore Ministry of Education Tier 2 Grant (MOE-T2EP50222-0018). We acknowledge the support of Dieter Schwarz

**CONFLICT OF INTERESTS**

The authors declare that they have no conflict of interests.

**CONTRIBUTONS**

All authors discussed the results and contributed to the manuscript.

# Reference:

1  Shannon, C. E. Communication theory of secrecy systems. *The Bell system technical journal* **28**, 656-715 (1949).
2  Devetak, I. The private classical capacity and quantum capacity of a quantum channel. *IEEE Transactions on Information Theory* **51**, 44-55 (2005).
3  Cai, N., Winter, A. & Yeung, R. W. Quantum privacy and quantum wiretap channels. *problems of information transmission* **40**, 318-336 (2004).
4  Wu, J., Long, G.-L. & Hayashi, M. Quantum secure direct communication with private dense coding using a general preshared quantum state. *Physical Review Applied* **17**, 064011 (2022).
5  Bennett, C. H. & Brassard, G. Quantum cryptography: Public key distribution and coin tossing. *Theoretical computer science* **560**, 7-11 (2014).
6  Renner, R., Gisin, N. & Kraus, B. Information-theoretic security proof for quantum-key-distribution protocols. *Physical Review A* **72**, 012332 (2005).
7  Bernstein, D. J. & Lange, T. Post-quantum cryptography. *Nature* **549**, 188-194 (2017).
8  Shor, P. W. Polynomial-time algorithms for prime factorization and discrete logarithms on a quantum computer. *SIAM review* **41**, 303-332 (1999).
9  Mehic, M. *et al.* Quantum key distribution: a networking perspective. *ACM Computing Surveys (CSUR)* **53**, 1-41 (2020).
10  Awschalom, D. D. *et al.* A Roadmap for Quantum Interconnects. Medium: ED; Size: 48 p. (United States, 2022).
11  Wehner, S., Elkouss, D. & Hanson, R. Quantum internet: A vision for the road ahead. *Science* **362**, eaam9288 (2018).
12  Bennett, C. H. Quantum cryptography using any two nonorthogonal states. *Physical review letters* **68**, 3121 (1992).
13  Vallone, G. *et al.* Free-space quantum key distribution by rotation-invariant twisted photons. *Physical review letters* **113**, 060503 (2014).
14  Inoue, K., Waks, E. & Yamamoto, Y. Differential phase shift quantum key distribution. *Physical review letters* **89**, 037902 (2002).
15  Stucki, D., Brunner, N., Gisin, N., Scarani, V. & Zbinden, H. Fast and simple one-way quantum key distribution. *Applied Physics Letters* **87** (2005).
16  Sasaki, T., Yamamoto, Y. & Koashi, M. Practical quantum key distribution protocol without monitoring signal disturbance. *Nature* **509**, 475-478 (2014).
17  Scarani, V. *et al.* The security of practical quantum key distribution. *Reviews of modern physics* **81**, 1301-1350 (2009).
18  Moroder, T. *et al.* Security of distributed-phase-reference quantum key distribution. *Physical review letters* **109**, 260501 (2012).
19  Ralph, T. C. Continuous variable quantum cryptography. *Physical Review A* **61**, 010303 (1999).
20  Reid, M. D. Quantum cryptography with a predetermined key, using continuous-variable Einstein-Podolsky-Rosen correlations. *Physical Review A* **62**, 062308 (2000).
21  Weedbrook, C. *et al.* Quantum cryptography without switching. *Physical review letters* **93**, 170504 (2004).
22  Grosshans, F. & Grangier, P. Continuous variable quantum cryptography using coherent states. *Physical review letters* **88**, 057902 (2002).
23  Grosshans, F. *et al.* Quantum key distribution using gaussian-modulated coherent states. *Nature* **421**, 238-241 (2003).

24  Hillery, M. Quantum cryptography with squeezed states. *Physical Review A* **61**, 022309 (2000).

25  Cerf, N. J., Levy, M. & Van Assche, G. Quantum distribution of Gaussian keys using squeezed states. *Physical Review A* **63**, 052311 (2001).

26  Weedbrook, C. *et al.* Gaussian quantum information. *Reviews of Modern Physics* **84**, 621-669 (2012).

27  Weedbrook, C., Pirandola, S., Lloyd, S. & Ralph, T. C. Quantum cryptography approaching the classical limit. *Physical review letters* **105**, 110501 (2010).

28  Silberhorn, C., Ralph, T. C., Lütkenhaus, N. & Leuchs, G. Continuous variable quantum cryptography: Beating the 3 dB loss limit. *Physical review letters* **89**, 167901 (2002).

29  Usenko, V. C. & Grosshans, F. Unidimensional continuous-variable quantum key distribution. *Physical Review A* **92**, 062337 (2015).

30  Albota, M. A. & Wong, F. N. Efficient single-photon counting at 1.55 μm by means of frequency upconversion. *Optics letters* **29**, 1449-1451 (2004).

31  Yuan, Z., Kardynal, B., Sharpe, A. & Shields, A. High speed single photon detection in the near infrared. *Applied Physics Letters* **91** (2007).

32  Gol'Tsman, G. *et al.* Picosecond superconducting single-photon optical detector. *Applied physics letters* **79**, 705-707 (2001).

33  Itzler, M. A. *et al.* Advances in InGaAsP-based avalanche diode single photon detectors. *Journal of Modern Optics* **58**, 174-200 (2011).

34  Ceccarelli, F. *et al.* Recent advances and future perspectives of single-photon avalanche diodes for quantum photonics applications. *Advanced Quantum Technologies* **4**, 2000102 (2021).

35  Ghioni, M., Gulinatti, A., Rech, I., Zappa, F. & Cova, S. Progress in silicon single-photon avalanche diodes. *IEEE Journal of selected topics in quantum electronics* **13**, 852-862 (2007).

36  Zappa, F., Lotito, A., Giudice, A. C., Cova, S. & Ghioni, M. Monolithic active-quenching and active-reset circuit for single-photon avalanche detectors. *IEEE Journal of Solid-State Circuits* **38**, 1298-1301 (2003).

37  Gallivanoni, A., Rech, I. & Ghioni, M. Progress in quenching circuits for single photon avalanche diodes. *IEEE Transactions on nuclear science* **57**, 3815-3826 (2010).

38  Acerbi, F., Anti, M., Tosi, A. & Zappa, F. Design criteria for InGaAs/InP single-photon avalanche diode. *IEEE Photonics Journal* **5**, 6800209-6800209 (2013).

39  Chang, J. *et al.* Detecting telecom single photons with 99.5− 2.07+ 0.5% system detection efficiency and high time resolution. *APL Photonics* **6** (2021).

40  Caloz, M. *et al.* High-detection efficiency and low-timing jitter with amorphous superconducting nanowire single-photon detectors. *Applied Physics Letters* **112** (2018).

41  Hu, P. *et al.* Detecting single infrared photons toward optimal system detection efficiency. *Optics Express* **28**, 36884-36891 (2020).

42  Chiles, J. *et al.* New constraints on dark photon dark matter with superconducting nanowire detectors in an optical haloscope. *Physical Review Letters* **128**, 231802 (2022).

43  Reddy, D. V., Nerem, R. R., Nam, S. W., Mirin, R. P. & Verma, V. B. Superconducting nanowire single-photon detectors with 98% system detection efficiency at 1550 nm. *Optica* **7**, 1649-1653 (2020).

44  You, L. Superconducting nanowire single-photon detectors for quantum information. *Nanophotonics* **9**, 2673-2692 (2020).

45  Renema, J. *et al.* Experimental test of theories of the detection mechanism in a nanowire superconducting single photon detector. *Physical review letters* **112**, 117604 (2014).

46  Qi, R., Zhang, H., Gao, J., Yin, L. & Long, G.-L. Loophole-free plug-and-play quantum key distribution. *New Journal of Physics* **23**, 063058 (2021).

47  Zhang, H. *et al.* Noise-reducing Quantum Key Distribution. *Reports on Progress in Physics* **88**, 016001 (2025).
48  Chen, J.-P. *et al.* Twin-field quantum key distribution over a 511 km optical fibre linking two distant metropolitan areas. *Nature Photonics* **15**, 570-575 (2021).
49  Avesani, M. *et al.* Full daylight quantum-key-distribution at 1550 nm enabled by integrated silicon photonics. *npj Quantum Information* **7**, 93 (2021).
50  Scriminich, A. *et al.* Optimal design and performance evaluation of free-space quantum key distribution systems. *Quantum Science and Technology* **7**, 045029 (2022).
51  Schmitt-Manderbach, T. *et al.* Experimental demonstration of free-space decoy-state quantum key distribution over 144 km. *Physical Review Letters* **98**, 010504 (2007).
52  Kržič, A. *et al.* Towards metropolitan free-space quantum networks. *npj Quantum Information* **9**, 95 (2023).
53  Tian, X.-H. *et al.* Experimental Demonstration of Drone-Based Quantum Key Distribution. *Physical Review Letters* **133**, 200801 (2024).
54  Liao, S.-K. *et al.* Satellite-to-ground quantum key distribution. *Nature* **549**, 43-47 (2017).
55  Liao, S.-K. *et al.* Long-distance free-space quantum key distribution in daylight towards inter-satellite communication. *Nature Photonics* **11**, 509-513 (2017).
56  Kundu, N. K., Dash, S. P., Mckay, M. R. & Mallik, R. K. Channel estimation and secret key rate analysis of MIMO terahertz quantum key distribution. *IEEE Transactions on Communications* **70**, 3350-3363 (2022).
57  Pirandola, S., Laurenza, R., Ottaviani, C. & Banchi, L. Fundamental limits of repeaterless quantum communications. *Nature communications* **8**, 15043 (2017).
58  Lucamarini, M., Yuan, Z. L., Dynes, J. F. & Shields, A. J. Overcoming the rate–distance limit of quantum key distribution without quantum repeaters. *Nature* **557**, 400-403 (2018).
59  Shi, B. *et al.* Splicing Hollow-Core Fiber with Standard Glass-Core Fiber with Ultralow Back-Reflection and Low Coupling Loss. *ACS photonics* **11**, 3288-3295 (2024).
60  Takesue, H. *et al.* Quantum key distribution over a 40-dB channel loss using superconducting single-photon detectors. *Nature photonics* **1**, 343-348 (2007).
61  Rey-Domínguez, J., Navarrete, Á., van Loock, P. & Curty, M. Hacking coherent-one-way quantum key distribution with present-day technology. *Quantum Science and Technology* **9**, 035044 (2024).
62  González-Payo, J., Trényi, R., Wang, W. & Curty, M. Upper security bounds for coherent-one-way quantum key distribution. *Physical Review Letters* **125**, 260510 (2020).
63  Grünenfelder, F., Boaron, A., Rusca, D., Martin, A. & Zbinden, H. Performance and security of 5 GHz repetition rate polarization-based quantum key distribution. *Applied Physics Letters* **117** (2020).
64  Boaron, A. *et al.* Secure quantum key distribution over 421 km of optical fiber. *Physical review letters* **121**, 190502 (2018).
65  Korzh, B. *et al.* Demonstration of sub-3 ps temporal resolution with a superconducting nanowire single-photon detector. *Nature Photonics* **14**, 250-255 (2020).
66  Bouchard, F., England, D., Bustard, P. J., Heshami, K. & Sussman, B. Quantum communication with ultrafast time-bin qubits. *PRX Quantum* **3**, 010332 (2022).
67  Wang, W. *et al.* Fully passive quantum key distribution. *Physical Review Letters* **130**, 220801 (2023).
68  Yuan, Z. *et al.* 10-Mb/s quantum key distribution. *Journal of Lightwave Technology* **36**, 3427-3433 (2018).
69  Zhang, W. *et al.* A 16-pixel interleaved superconducting nanowire single-photon detector array with a maximum count rate exceeding 1.5 GHz. *IEEE Transactions on Applied Superconductivity* **29**, 1-4 (2019).
70  Grünenfelder, F. *et al.* Fast single-photon detectors and real-time key distillation enable high secret-key-rate quantum key distribution systems. *Nature Photonics* **17**, 422-426 (2023).
71  Li, W. *et al.* High-rate quantum key distribution exceeding 110 Mb s−1. *Nature Photonics* **17**, 416-421 (2023).

| | |
|---|---|
| 72 | Cavaliere, F., Prati, E., Poti, L., Muhammad, I. & Catuogno, T. Secure quantum communication technologies and systems: From labs to markets. *Quantum Reports* **2**, 80-106 (2020). |
| 73 | Dynes, J. *et al.* Cambridge quantum network. *npj Quantum Information* **5**, 101 (2019). |
| 74 | Dou, T. *et al.* Coexistence of 11 Tbps (110× 100 Gbps) classical optical communication and quantum key distribution based on single-mode fiber. *Optics Express* **32**, 28356-28369 (2024). |
| 75 | Kumar, R., Qin, H. & Alléaume, R. Coexistence of continuous variable QKD with intense DWDM classical channels. *New Journal of Physics* **17**, 043027 (2015). |
| 76 | Mao, Y. *et al.* Integrating quantum key distribution with classical communications in backbone fiber network. *Optics express* **26**, 6010-6020 (2018). |
| 77 | Wang, B.-X. *et al.* Long-distance transmission of quantum key distribution coexisting with classical optical communication over a weakly-coupled few-mode fiber. *Optics express* **28**, 12558-12565 (2020). |
| 78 | Simmons, J. M. *Optical Network Design and Planning.* (Springer International Publishing, 2014). |
| 79 | Cao, Y. *et al.* The evolution of quantum key distribution networks: On the road to the qinternet. *IEEE Communications Surveys & Tutorials* **24**, 839-894 (2022). |
| 80 | Bunandar, D. *et al.* Metropolitan quantum key distribution with silicon photonics. *Physical Review X* **8**, 021009 (2018). |
| 81 | Luo, W. *et al.* Recent progress in quantum photonic chips for quantum communication and internet. *Light: Science & Applications* **12**, 175 (2023). |
| 82 | Zhang, G.-W. *et al.* Polarization-insensitive interferometer based on a hybrid integrated planar light-wave circuit. *Photonics Research* **9**, 2176-2181 (2021). |
| 83 | Renaud, D. *et al.* Sub-1 Volt and high-bandwidth visible to near-infrared electro-optic modulators. *Nature Communications* **14**, 1496 (2023). |
| 84 | Paraïso, T. K. *et al.* A modulator-free quantum key distribution transmitter chip. *npj Quantum Information* **5**, 42 (2019). |
| 85 | Sibson, P. *et al.* Chip-based quantum key distribution. *Nature communications* **8**, 13984 (2017). |
| 86 | Najafi, F. *et al.* On-chip detection of non-classical light by scalable integration of single-photon detectors. *Nature communications* **6**, 5873 (2015). |
| 87 | Martinez, N. J. *et al.* Single photon detection in a waveguide-coupled Ge-on-Si lateral avalanche photodiode. *Optics express* **25**, 16130-16139 (2017). |
| 88 | Zhang, G. *et al.* An integrated silicon photonic chip platform for continuous-variable quantum key distribution. *Nature Photonics* **13**, 839-842 (2019). |
| 89 | Bruynsteen, C., Vanhoecke, M., Bauwelinck, J. & Yin, X. Integrated balanced homodyne photonic–electronic detector for beyond 20 GHz shot-noise-limited measurements. *Optica* **8**, 1146-1152 (2021). |
| 90 | Ma, R. *et al.* Disorder enhanced relative intrinsic detection efficiency in NbTiN superconducting nanowire single photon detectors at high temperature. *Applied Physics Letters* **124**, 072601 (2024). |
| 91 | Gemmell, N. R. *et al.* A miniaturized 4 K platform for superconducting infrared photon counting detectors. *Superconductor Science and Technology* **30**, 11LT01 (2017). |
| 92 | Charaev, I. *et al.* Single-photon detection using large-scale high-temperature MgB2 sensors at 20 K. *Nature Communications* **15**, 3973 (2024). |
| 93 | Na, N. *et al.* Room temperature operation of germanium–silicon single-photon avalanche diode. *Nature* **627**, 295-300 (2024). |
| 94 | Jouguet, P., Kunz-Jacques, S. & Diamanti, E. Preventing calibration attacks on the local oscillator in continuous-variable quantum key distribution. *Physical Review A* **87**, 062313 (2013). |
| 95 | Mao, Y. *et al.* Hidden-Markov-model-based calibration-attack recognition for continuous-variable quantum |

key distribution. *Physical Review A* **101**, 062320 (2020).

96    Qin, H., Kumar, R. & Alléaume, R. Quantum hacking: Saturation attack on practical continuous-variable quantum key distribution. *Physical Review A* **94**, 012325 (2016).

97    Huang, J.-Z. et al. Quantum hacking of a continuous-variable quantum-key-distribution system using a wavelength attack. *Physical Review A* **87**, 062329 (2013).

98    Huang, A. et al. Laser-seeding attack in quantum key distribution. *Physical Review Applied* **12**, 064043 (2019).

99    Tang, Y.-L. et al. Source attack of decoy-state quantum key distribution using phase information. *Physical Review A* **88**, 022308 (2013).

100   Xu, F., Qi, B. & Lo, H.-K. Experimental demonstration of phase-remapping attack in a practical quantum key distribution system. *New Journal of Physics* **12**, 113026 (2010).

101   Fung, C.-H. F., Qi, B., Tamaki, K. & Lo, H.-K. Phase-remapping attack in practical quantum-key-distribution systems. *Physical Review A* **75**, 032314 (2007).

102   Qin, H., Kumar, R., Makarov, V. & Alléaume, R. Homodyne-detector-blinding attack in continuous-variable quantum key distribution. *Physical Review A* **98**, 012312 (2018).

103   Yuan, Z., Dynes, J. F. & Shields, A. J. Avoiding the blinding attack in QKD. *Nature Photonics* **4**, 800-801 (2010).

104   Gerhardt, I. et al. Full-field implementation of a perfect eavesdropper on a quantum cryptography system. *Nature communications* **2**, 349 (2011).

105   Ma, X.-C., Sun, S.-H., Jiang, M.-S. & Liang, L.-M. Local oscillator fluctuation opens a loophole for Eve in practical continuous-variable quantum-key-distribution systems. *Physical Review A* **88**, 022339 (2013).

106   Ding, C. et al. Machine-learning-based detection for quantum hacking attacks on continuous-variable quantum-key-distribution systems. *Physical Review A* **107**, 062422 (2023).

107   Lucamarini, M. et al. Practical security bounds against the trojan-horse attack in quantum key distribution. *Physical Review X* **5**, 031030 (2015).

108   Lo, H.-K., Curty, M. & Qi, B. Measurement-device-independent quantum key distribution. *Physical review letters* **108**, 130503 (2012).

109   Braunstein, S. L. & Pirandola, S. Side-channel-free quantum key distribution. *Physical review letters* **108**, 130502 (2012).

110   Wang, W., Xu, F. & Lo, H.-K. Asymmetric protocols for scalable high-rate measurement-device-independent quantum key distribution networks. *Physical Review X* **9**, 041012 (2019).

111   Tang, Y.-L. et al. Measurement-device-independent quantum key distribution over untrustful metropolitan network. *Physical Review X* **6**, 011024 (2016).

112   Zapatero, V. et al. Advances in device-independent quantum key distribution. *npj quantum information* **9**, 10 (2023).

113   Arnon-Friedman, R., Dupuis, F., Fawzi, O., Renner, R. & Vidick, T. Practical device-independent quantum cryptography via entropy accumulation. *Nature communications* **9**, 459 (2018).

114   Clauser, J. F., Horne, M. A., Shimony, A. & Holt, R. A. Proposed experiment to test local hidden-variable theories. *Physical review letters* **23**, 880 (1969).

115   Hensen, B. et al. Loophole-free Bell inequality violation using electron spins separated by 1.3 kilometres. *Nature* **526**, 682-686 (2015).

116   Liu, W.-Z. et al. Toward a photonic demonstration of device-independent quantum key distribution. *Physical Review Letters* **129**, 050502 (2022).

117   Tan, E. Y.-Z. & Wolf, R. Entropy bounds for device-independent quantum key distribution with local Bell test. *Physical Review Letters* **133**, 120803 (2024).

118   Steffinlongo, A. et al. Long-distance device-independent quantum key distribution using single-photon entanglement. *arXiv preprint arXiv:2409.17075* (2024).
119  Gisin, N., Pironio, S. & Sangouard, N. Proposal for Implementing Device-Independent Quantum Key Distribution Based on a Heralded Qubit Amplifier. *Physical review letters* **105**, 070501 (2010).
120  Zhang, W. *et al.* A device-independent quantum key distribution system for distant users. *Nature* **607**, 687-691 (2022).
121  Nadlinger, D. P. *et al.* Experimental quantum key distribution certified by Bell's theorem. *Nature* **607**, 682-686 (2022).
122  Townsend, P. D., Rarity, J. & Tapster, P. Single photon interference in 10 km long optical fibre interferometer. *Electronics Letters* **7**, 634-635 (1993).
123  Stucki, D., Gisin, N., Guinnard, O., Ribordy, G. & Zbinden, H. Quantum key distribution over 67 km with a plug&play system. *New Journal of Physics* **4**, 41 (2002).
124  Lo, H.-K., Ma, X. & Chen, K. Decoy state quantum key distribution. *Physical review letters* **94**, 230504 (2005).
125  Hwang, W.-Y. Quantum key distribution with high loss: toward global secure communication. *Physical review letters* **91**, 057901 (2003).
126  Wang, X.-B. Beating the photon-number-splitting attack in practical quantum cryptography. *Physical review letters* **94**, 230503 (2005).
127  Peng, C.-Z. *et al.* Experimental long-distance decoy-state quantum key distribution based on polarization encoding. *Physical review letters* **98**, 010505 (2007).
128  Stucki, D. *et al.* High rate, long-distance quantum key distribution over 250 km of ultra low loss fibres. *New Journal of Physics* **11**, 075003 (2009).
129  Yin, H.-L. *et al.* Measurement-device-independent quantum key distribution over a 404 km optical fiber. *Physical review letters* **117**, 190501 (2016).
130  Liao, S.-K. *et al.* Satellite-relayed intercontinental quantum network. *Physical review letters* **120**, 030501 (2018).
131  Chen, J.-P. *et al.* Sending-or-not-sending with independent lasers: Secure twin-field quantum key distribution over 509 km. *Physical review letters* **124**, 070501 (2020).
132  Wang, S. *et al.* Twin-field quantum key distribution over 830-km fibre. *Nature photonics* **16**, 154-161 (2022).
133  Zeng, P., Zhou, H., Wu, W. & Ma, X. Mode-pairing quantum key distribution. *Nature Communications* **13**, 3903 (2022).
134  Liu, Y. *et al.* Experimental twin-field quantum key distribution over 1000 km fiber distance. *Physical Review Letters* **130**, 210801 (2023).
135  Cao, Y. *et al.* From single-protocol to large-scale multi-protocol quantum networks. *IEEE Network* **36**, 14-22 (2022).
136  Liu, J.-L. *et al.* Creation of memory–memory entanglement in a metropolitan quantum network. *Nature* **629**, 579-585, doi:10.1038/s41586-024-07308-0 (2024).
137  Peev, M. *et al.* The SECOQC quantum key distribution network in Vienna. *New Journal of Physics* **11**, 075001 (2009).
138  Yang, C., Zhang, H. & Su, J. The QKD network: Model and routing scheme. *Journal of Modern Optics* **64**, 2350-2362 (2017).
139  Cao, Y. *et al.* Hybrid trusted/untrusted relay-based quantum key distribution over optical backbone networks. *IEEE Journal on Selected Areas in Communications* **39**, 2701-2718 (2021).
140  Long, G.-L. *et al.* An evolutionary pathway for the quantum internet relying on secure classical repeaters. *IEEE Network* **36**, 82-88 (2022).
141  Garms, L. *et al.* Experimental Integration of Quantum Key Distribution and Post-Quantum Cryptography in a Hybrid Quantum-Safe Cryptosystem. *Advanced Quantum Technologies* **7**, 2300304,

doi:https://doi.org/10.1002/qute.202300304 (2024).

142  Long, G.-L. & Liu, X.-S. Theoretically efficient high-capacity quantum-key-distribution scheme. *Physical Review A* **65**, 032302 (2002).

143  Wang, M. & Long, G.-L. Lattice-based access authentication scheme for quantum communication networks. *Science China Information Sciences* **67**, 222501 (2024).

144  Xie, Y.-M. *et al.* Breaking the rate-loss bound of quantum key distribution with asynchronous two-photon interference. *PRX Quantum* **3**, 020315 (2022).

145  Wang, S. *et al.* Beating the fundamental rate-distance limit in a proof-of-principle quantum key distribution system. *Physical Review X* **9**, 021046 (2019).

146  Minder, M. *et al.* Experimental quantum key distribution beyond the repeaterless secret key capacity. *Nature Photonics* **13**, 334-338 (2019).

147  Liu, Y. *et al.* Experimental twin-field quantum key distribution through sending or not sending. *Physical Review Letters* **123**, 100505 (2019).

148  Zhong, X., Hu, J., Curty, M., Qian, L. & Lo, H.-K. Proof-of-principle experimental demonstration of twin-field type quantum key distribution. *Physical Review Letters* **123**, 100506 (2019).

149  Fang, X.-T. *et al.* Implementation of quantum key distribution surpassing the linear rate-transmittance bound. *Nature Photonics* **14**, 422-425 (2020).

150  Liu, H. *et al.* Field test of twin-field quantum key distribution through sending-or-not-sending over 428 km. *Physical Review Letters* **126**, 250502 (2021).

151  Zhu, H.-T. *et al.* Experimental mode-pairing measurement-device-independent quantum key distribution without global phase locking. *Physical Review Letters* **130**, 030801 (2023).

152  Zhou, L. *et al.* Experimental quantum communication overcomes the rate-loss limit without global phase tracking. *Physical Review Letters* **130**, 250801 (2023).

153  Pirandola, S. End-to-end capacities of a quantum communication network. *Communications Physics* **2**, 51 (2019).

154  Lu, C.-Y., Cao, Y., Peng, C.-Z. & Pan, J.-W. Micius quantum experiments in space. *Reviews of Modern Physics* **94**, 035001 (2022).

155  Bedington, R., Arrazola, J. M. & Ling, A. Progress in satellite quantum key distribution. *npj Quantum Information* **3**, 30 (2017).

156  Hosseinidehaj, N., Babar, Z., Malaney, R., Ng, S. X. & Hanzo, L. Satellite-based continuous-variable quantum communications: State-of-the-art and a predictive outlook. *IEEE Communications Surveys & Tutorials* **21**, 881-919 (2018).

157  Villoresi, P. *et al.* Experimental verification of the feasibility of a quantum channel between space and Earth. *New Journal of Physics* **10**, 033038 (2008).

158  Nauerth, S. *et al.* Air-to-ground quantum communication. *Nature Photonics* **7**, 382-386 (2013).

159  Wang, J.-Y. *et al.* Direct and full-scale experimental verifications towards ground–satellite quantum key distribution. *Nature Photonics* **7**, 387-393 (2013).

160  Elliott, C. *et al.* in *Quantum Information and computation III.*  138-149 (SPIE).

161  Chen, T.-Y. *et al.* Field test of a practical secure communication network with decoy-state quantum cryptography. *Optics express* **17**, 6540-6549 (2009).

162  Sasaki, M. *et al.* Field test of quantum key distribution in the Tokyo QKD Network. *Optics express* **19**, 10387-10409 (2011).

163  Chen, T.-Y. *et al.* Metropolitan all-pass and inter-city quantum communication network. *Optics express* **18**, 27217-27225 (2010).

164  Wang, S. *et al.* Field test of wavelength-saving quantum key distribution network. *Optics letters* **35**, 2454-2456 (2010).

165  Stucki, D. *et al.* Long-term performance of the SwissQuantum quantum key distribution network in a field environment. *New Journal of Physics* **13**, 123001 (2011).

166  Chen, Y.-A. *et al.* An integrated space-to-ground quantum communication network over 4,600 kilometres. *Nature* **589**, 214-219 (2021).

167  Cao, Y. *et al.* Long-distance free-space measurement-device-independent quantum key distribution. *Physical Review Letters* **125**, 260503 (2020).

168  Ji, L. *et al.* Towards quantum communications in free-space seawater. *Optics Express* **25**, 19795-19806 (2017).

169  Chen, T.-Y. *et al.* Implementation of a 46-node quantum metropolitan area network. *npj Quantum Information* **7**, 134 (2021).

170  Liu, H. *et al.* Experimental demonstration of high-rate measurement-device-independent quantum key distribution over asymmetric channels. *Physical review letters* **122**, 160501 (2019).

171  Zhong, X., Wang, W., Qian, L. & Lo, H.-K. Proof-of-principle experimental demonstration of twin-field quantum key distribution over optical channels with asymmetric losses. *npj Quantum Information* **7**, 8 (2021).

172  Zhu, H.-T. *et al.* Field test of mode-pairing quantum key distribution. *Optica* **11**, 883-888 (2024).

173  Chen, J.-P. *et al.* Twin-Field Quantum Key Distribution with Local Frequency Reference. *Physical Review Letters* **132**, 260802, doi:10.1103/PhysRevLett.132.260802 (2024).

174  Bennett, C. H., Bessette, F., Brassard, G., Salvail, L. & Smolin, J. Experimental quantum cryptography. *Journal of cryptology* **5**, 3-28 (1992).

175  Muller, A., Breguet, J. & Gisin, N. Experimental demonstration of quantum cryptography using polarized photons in optical fibre over more than 1 km. *Europhysics Letters* **23**, 383 (1993).

176  Li, C. *et al.* Practical High-efficiency SNSPD Balanced High Speed and Low Jitter on Filter Circuit. *IEEE Transactions on Applied Superconductivity* **34**, 1-7 (2024).

177  Zhang, C. *et al.* Experimental side-channel-secure quantum key distribution. *Physical Review Letters* **128**, 190503 (2022).

178  Li, Y. *et al.* Microsatellite-based real-time quantum key distribution. *arXiv preprint arXiv:2408.10994* (2024).

179  Clark, D. D., Pogran, K. T. & Reed, D. P. An introduction to local area networks. *Proceedings of the IEEE* **66**, 1497-1517 (1978).

180  Sze, D. A metropolitan area network. *IEEE Journal on Selected Areas in Communications* **3**, 815-824 (1985).

181  Pan, D. *et al.* The evolution of quantum secure direct communication: on the road to the qinternet. *IEEE Communications Surveys & Tutorials* **26**, 1898 - 1949 (2024).

182  Makarov, V. *et al.* Preparing a commercial quantum key distribution system for certification against implementation loopholes. *Physical Review Applied* **22**, 044076 (2024).

183  Länger, T. & Lenhart, G. Standardization of quantum key distribution and the ETSI standardization initiative ISG-QKD. *New Journal of Physics* **11**, 055051 (2009).